\documentclass[a4paper,11pt]{article}
\pdfoutput=1 

\usepackage{jcappub,wrapfig,verbatim,amsthm,graphicx,hyperref,enumerate,enumitem} 
                     
\usepackage[utf8]{inputenc}

\title{\boldmath A Cosmological Underdensity Does Not Solve the Hubble Tension}

\author[a]{Sveva Castello,}
\affiliation[a]{Department of Physics and Astronomy, Uppsala University, SE 751 20 Uppsala, Sweden}

\author[b]{Marcus Högås}
\affiliation[b]{The Oskar Klein Centre, Department of Physics, Stockholm University, Stockholm SE 106 91, Sweden}

\author[b]{and Edvard Mörtsell}

\emailAdd{sveva.castello@unige.ch, marcus.hogas@fysik.su.se, edvard@fysik.su.se}

\abstract{A potential solution to the Hubble tension is the hypothesis that the Milky Way is located near the center of a matter underdensity. We model this scenario through the Lemaître--Tolman--Bondi formalism with the inclusion of a cosmological constant ($\Lambda$LTB) and consider a generalized Gaussian parametrization for the matter density profile. We constrain the underdensity and the background cosmology with a combination of data sets: the Pantheon Sample of type Ia supernovae (both the full catalogue and a redshift-binned version of it), a collection of baryon acoustic oscillations data points and the distance priors extracted from the latest \textit{Planck} data release. The analysis with the binned supernovae suggests a preference for a $-13 \%$ density drop with a size of approximately 300 Mpc, interestingly matching the prediction for the so-called KBC void already identified on the basis of independent analyses using galaxy distributions. The constraints obtained with the full Pantheon Sample are instead compatible with a homogeneous cosmology and we interpret this radically different result as a cautionary tale about the potential bias introduced by employing a binned supernova data set. We quantify the level of improvement on the Hubble tension by analyzing the constraints on the B-band absolute magnitude of the supernovae, which provides the calibration for the local measurements of $H_0$. Since no significant difference is observed with respect to an analogous fit performed with a standard $\Lambda$CDM cosmology, we conclude that the potential presence of a local underdensity does not resolve the tension and does not significantly degrade current supernova constraints on $H_0$.}

\begin{document}
\maketitle
\flushbottom

\section{Introduction} \label{sec:introduction}
In recent years, the increasingly precise measurements of the Hubble constant, $H_0$, have revealed a potential schism between the constraints arising from observations at late and early cosmological times, respectively corresponding to small and large distances from us. This so-called Hubble tension is particularly evident when considering the SH0ES team's constraint of $H_0 = (73.2 \pm 1.3)$ km s$^{-1}$ Mpc$^{-1}$, obtained by calibrating type Ia supernovae (SNe Ia) distances up to $z \simeq 0.15$ through Cepheid variable stars \cite{Riess:2020fzl}. This implies a 4.2$\sigma$ discrepancy with the early-time \textit{Planck} satellite's result of $H_0 = (67.4 \pm 0.5)$ km s$^{-1}$ Mpc$^{-1}$, based upon the distance measurement to the cosmic microwave background (CMB) decoupling surface at $z \simeq 1090$  \cite{Planck:2018vyg}. Other late-time measurements obtained with different techniques corroborate the tension (see e.g. \cite{Wong:2019kwg, Pesce:2020xfe, Kourkchi:2020iyz, Schombert:2020pxm,Blakeslee:2021rqi}), while the calibration of the local distance ladder via the tip of the red giant branch method (TRGB) yields a value of $H_0 = (69.8 \pm 1.6)$ km s$^{-1}$ Mpc$^{-1}$ \cite{Freedman:2021ahq}, which is interestingly located right between the SH0ES and the \textit{Planck} constraints.

The Hubble tension has sparked much interest and excitement in the cosmology community and several explanations have been proposed \cite{DiValentino:2021izs, Efstathiou:2021ocp}. A solution that does not require the introduction of new physics consists in postulating the presence of a matter underdensity in the local Universe, i.e. a cosmic void. This would result in a faster local expansion, leading to a larger local value of $H_0$ and thus potentially alleviating the tension. The existence of such a ``local hole" was already suggested in previous studies based upon the reconstructed luminosity density and galaxy number counts in the local Universe. Most notably, Keenan \textit{et al.} \cite{Keenan:2013mfa} provided observational evidence for the so-called Keenan--Barger--Cowie (KBC) void, corresponding to a $\simeq -30\%$ underdensity with a radius of around 300 Mpc roughly centered on the Milky Way. Other authors \cite{Frith:2003tb, Whitbourn:2013mwa, Bohringer:2019tyj, Wong:2021fvu} found similar indications on the scale of a few hundreds Mpc, with slight variations in size and depth with respect to the KBC void. 

These results were questioned by Kenworthy \textit{et al.} \cite{Kenworthy:2019qwq} (hereafter K19), who employed the Pantheon Sample of SNe Ia \cite{Scolnic:2017caz} to show that there is no evidence of abrupt density variations at low redshifts in the supernova data. Moreover, N-body simulations \cite{Odderskov:2014hqa, Wu:2017fpr} suggested that, even if the presence of a local void is not excluded, the depth required to reconcile the Hubble tension is not likely to be attained in the standard paradigm for cosmology, the $\Lambda$CDM model. The hypothesis of homogeneity in the local Universe was also recently tested by Camarena \textit{et al.} \cite{Camarena:2021mjr} (hereafter C21) with a combination of cosmological probes: the 2018 \textit{Planck} data \cite{Planck:2018vyg}, the Pantheon Sample of SNe Ia \cite{Scolnic:2017caz} (including a local prior on the SNe absolute magnitude \cite{Camarena:2021jlr}, and therefore on $H_0$), cosmic chronometers \cite{Moresco:2015cya}, baryon acoustic oscillations (BAO) distances \cite{Beutler2011, Ross:2014qpa, BOSS:2016wmc}, the Compton $y$-distortion \cite{Fixsen:1996nj} and the kinetic Sunyaev-Zeldovich effect \cite{SPT:2020psp}. Their work, which appeared while this manuscript was being finalized, measures the deviation from homogeneity to be a mere $\sim 1 \%$ effect on scales $\gtrsim 100$ Mpc and finds that this does not impact the constraints on the cosmological parameters. On the other hand, it has been suggested \cite{Lombriser:2019ahl} that a matter density fluctuation on a 40 Mpc scale could provide a borderline consistency between the local SNe Ia measurement of $H_0$ and the \textit{Planck} one. 

This scenario is the starting point for this work\footnote{This paper is based upon the master's thesis \cite{Castello1571650} by Sveva Castello, available at \url{http://urn.kb.se/resolve?urn=urn:nbn:se:uu:diva-446960}.}, which aims at investigating the ``void solution" to the Hubble tension by constraining the properties of the local matter density profile. We model a generic matter density fluctuation within the standard framework of the Lemaître--Tolman--Bondi metric \cite{lemaitre1997expanding, tolman1934effect, bondi1947spherically} in the presence of a cosmological constant ($\Lambda$LTB) with a rigorous and physically meaningful set of boundary conditions. We then perform a fit to a combination of up-to-date data sets: the Pantheon Sample of SNe Ia \cite{Scolnic:2017caz}, a collection of BAO data points from different galaxy surveys \cite{Beutler2011, Ross:2014qpa, BOSS:2016wmc, Bautista:2017wwp, Zhao:2018gvb} and the distance priors extracted from the latest \textit{Planck} data release \cite{Chen:2018dbv}. In particular, we test the constraining power of the SNe data by carrying out the analysis both with the full Pantheon Sample and with a redshift-binned version of it. In contrast to C21, we do not employ cosmic chronometers, since it was shown that they cannot provide a high degree of precision in measuring the Hubble parameter \cite{Kjerrgren:2021zuo}.

Our modelling of a generic inhomogeneity generalizes the approach adopted by K19, who searched for observational signatures of specific void sizes and depths. Moreover, instead of solely employing SNe Ia data, we extend the analyses by K19 and \cite{Lukovic:2019ryg, Cai:2020tpy} by considering a more complete set of observations covering a broader redshift range. This allows to constrain the properties of the local matter density profile with increased precision. Interestingly, we find that the best-fit values for the parameters describing the inhomogeneity are dependent on the redshift binning of the Pantheon Sample, which can be interpreted as a cautionary tale for the use of a compressed supernova data set. We also argue that our procedure provides an immediate way to quantify the level of improvement in the Hubble tension, yielding a strong indication that a local underdensity cannot reconcile the discrepancy. 

\paragraph{Notation.} We adopt units where the speed of light $c$ is set to 1 and indicate partial derivatives with respect to the proper time $t$ and the radial coordinate $r$ with dots and primes respectively. The label ``asy" denotes asymptotic quantities evaluated at $r \rightarrow \infty$.

\section{Lemaître--Tolman--Bondi models with a cosmological constant} \label{sec:LTB}
\subsection{General framework} \label{sec:general_framework}
LTB solutions to Einstein's field equations of general relativity \citep{lemaitre1997expanding, tolman1934effect, bondi1947spherically} describe an inhomogeneous and isotropic cosmology, providing a suitable formalism to model a spherically symmetric density fluctuation with a matter distribution that varies along the radial coordinate. We refer to \cite{Castello1571650} for a complete discussion and derivation of the LTB set-up and only present the key results below. 

The standard form of the LTB metric is
\begin{equation} \label{eq:LTB_metric}
    \mathrm{d} s^2 = -\mathrm{d} t^2 + \frac{R'^2(t, r)}{1 - k(r)} \mathrm{d} r^2 + R^2 (t, r) \, (\mathrm{d} \theta^2 + \sin{\theta}^2 \mathrm{d} \phi^2),
\end{equation}
where $R(t, r)$ can be interpreted as a generalized scale factor depending both on the proper time $t$ and the radial coordinate $r$ and $k(r) < 1$ is a function related to the curvature of the hypersurfaces $t \equiv \mathrm{const}$. The Friedmann--Lemaître--Robertson--Walker (FLRW) metric can be recovered with the substitutions
\begin{equation} \label{eq:FLRW_from_LTB}
    R(t, r) \rightarrow a(t) r \, \, \, \mathrm{and} \, \, \, k(r) \rightarrow k r^2, 
\end{equation}
where $a(t)$ is the standard cosmological scale factor and $k$ is a constant describing the intrinsic spatial curvature of the metric.

While the original LTB formulation only includes curvature and matter in the form of pressureless dust, we consider the more general class of ``$\Lambda$LTB" models, which also incorporate the contribution of a cosmological constant $\Lambda$ with density $\rho_\Lambda = \mathrm{const}$. The equations of motion associated with the metric yield the generalized Friedmann equation,
\begin{equation} \label{eq:generalized_Friedmann}
     \frac{H_T^2(t, r)}{H_{T,0}^2(r)} = \Omega_{m,0}(r) \left(\frac{R_0(r)}{R(t, r)}\right)^3 + \Omega_{k,0}(r) \left(\frac{R_0(r)}{R(t, r)}\right)^2 + \Omega_\Lambda(r),
\end{equation}
where the subscript 0 indicates quantities evaluated at an arbitrary initial time $t_0(r)$. Here, we have defined generalized versions of the Hubble parameter and of the Hubble constant that include a radial dependence:
\begin{equation} \label{eq:LTB_Hubble}
    H_T(t, r) \equiv \frac{\dot{R}(t, r)}{R(t, r)} \implies H_{T,0}(r) \equiv \frac{\dot{R}_0(r)}{R_0(r)}.
\end{equation}
The subscript $T$ refers to the fact that these are ``transverse" quantities, i.e. corresponding to angular directions. Because of the radial dependence of the generalized scale factor, it is also possible to introduce a ``radial" Hubble parameter, 
\begin{equation} \label{eq:H_R}
    H_R(t, r) \equiv \dot{R}'(t, r) / R'(t, r).
\end{equation}

The density parameters evaluated at $t_0(r)$ satisfy the relation $\Omega_{m,0}(r) + \Omega_{k,0}(r) + \Omega_\Lambda(r) = 1$ and can be written as
\begin{equation} \label{eq:LTB_density_parameters}
    \Omega_{m,0}(r) \equiv \frac{\tilde{M}(r)}{H_0^2(r) R_0^3(r)}, \, \, \, \, \Omega_{k,0}(r) \equiv -\frac{k(r)}{H_0^2(r) R_0^2(r)} \, \, \, \mathrm{and} \, \, \, \Omega_\Lambda(r) \equiv \frac{\kappa \rho_\Lambda}{3 H_0^2(r)}. 
\end{equation}
Here, we have defined $\kappa = 8\pi G$, where $G$ is Newton's constant of gravitation, and we have introduced the LTB mass function 
\begin{equation} \label{eq:mass_fn}
    \tilde{M}(r) = \kappa \int^r_0 \mathrm{d} \tilde{r} \, R'(t, \tilde{r}) R^2(t, \tilde{r}) \rho_{m, 0}(\tilde{r}),
\end{equation}
where $\rho_{m, 0}(r)$ is the matter density at $t_0(r)$. In the following, we will fix the gauge freedom of $R(t,r)$ with the standard choice \cite{Enqvist:2007vb}
\begin{equation} \label{eq:gauge}
    R_0(r) \equiv R(t_0(r), r) = r,
\end{equation} 
simplifying Eqs. (\ref{eq:generalized_Friedmann})--(\ref{eq:mass_fn}).

The key difference with respect to a FLRW universe is that all cosmological parameters contain a radial dependence, incorporating inhomogeneities both in the expansion rate (through $H_{T, 0}(r)$) and in the matter distribution (through $\rho_{m,0}(r)$). As pointed out in \cite{Enqvist:2007vb}, these two functions consist of two physically distinct degrees of freedom, since they have independent boundary conditions, and must be specified separately in order to (numerically) solve the generalized Friedmann equation. We employ the set-up adopted in \cite{Hoscheit:2018nfl} and K19, which we implement with a more rigorous and physically motivated set of boundary conditions. 

First, we set $t_0$ to the elapsed time since the Big Bang (i.e. the age of the Universe) $t_{BB}$, which we assume to be the same at all radii, 
\begin{equation} \label{t_BB}
    t_0 \equiv t_{BB} = \mathrm{const.}
\end{equation}
Additionally, instead of directly referring to $\rho_{m,0}(r)$, we use the dimensionless matter density contrast $\delta(r)$,
\begin{equation} \label{eq:delta_def}
     \delta(r) \equiv \frac{\rho_{m, 0}(r) - \rho_{m,0}^\mathrm{asy}}{\rho_{m,0}^\mathrm{asy}}, \, \, \, \mathrm{with} \, \, \, \rho_{m,0}^\mathrm{asy} = \frac{3 \Omega_{m,0}^\mathrm{asy} \, (H_0^\mathrm{asy})^2}{\kappa},
\end{equation}
which can be interpreted as the matter density profile and therefore provides a more intuitive description than $\rho_{m,0}(r)$. We perform our main analysis with a generalized Gaussian ansatz, specified in Sec. \ref{sec:GG}. Additional results obtained with an Oppenheimer--Snyder (OS) ansatz \cite{OSpaper} are presented in Appendix \ref{app:OS}.

Together with the gauge choice in Eq. (\ref{eq:gauge}), the conditions on $t_0$ and $\delta(r)$ give the mass function 
\begin{equation}
    \tilde{M}(r) = \kappa \int^r_0 \mathrm{d} \tilde{r} \, \tilde{r}^2 \rho_{m, 0}(\tilde{r}) = 3 \Omega_{m,0}^\mathrm{asy} \, (H_0^\mathrm{asy})^2 \int^r_0 \mathrm{d} \tilde{r} \, \tilde{r}^2  (1 + \delta(\tilde{r})).
\end{equation}
We also adopt the standard assumption that the space-time is asymptotically flat, such that
\begin{equation}
    \Omega_\Lambda^\mathrm{asy} = 1 - \Omega_{m,0}^\mathrm{asy}.
\end{equation} 
This set-up yields the following set of relations:
\begin{equation} \label{eq:LTB_density_parameters_delta}
    \left\{
    \begin{aligned}
        & \Omega_{m,0}(r) = \frac{3 \Omega_{m,0}^\mathrm{asy}}{r^3} \, \left(\frac{H_0^\mathrm{asy}}{H_{T,0}(r)}\right)^2 \,\int^r_0 \mathrm{d} \tilde{r} \, \tilde{r}^2  (1 + \delta(\tilde{r})) \\
        & \Omega_\Lambda(r) = (1 - \Omega_{m,0}^\mathrm{asy}) \left(\frac{H_0^\mathrm{asy}}{H_{T,0}(r)}\right)^2 \\
        & \Omega_{k,0}(r) = 1 - \Omega_{m,0}(r) - \Omega_\Lambda(r),
    \end{aligned}
  \right.
\end{equation}
where, for a specified $\delta(r)$ and fixed values of $\Omega_{m,0}^\mathrm{asy}$ and $H_0^\mathrm{asy}$, the density parameters are given as functions of $H_{T,0}(r)$ only. The latter can be determined from rewriting the generalized Friedmann equation (\ref{eq:generalized_Friedmann}) as 
\begin{equation} \label{eq:H_0_r}
    \int_0^r \frac{\mathrm{d} R}{H_{T,0}(r) \sqrt{\Omega_{m,0}(r) \, r^3 R^{-1} + \Omega_{k,0}(r) \, r^2 + \Omega_{\Lambda}(r) \, R^2}} = t_{BB} = \mathrm{const.}
\end{equation}
We specify the age of the Universe $t_{BB}$ in the limit of asymptotic flatness, where the space-time reduces to the FLRW case with zero curvature,
\begin{equation} \label{eq:t_BB}
    t_{BB} = \int_0^1 \frac{\mathrm{d} a}{H_0^\mathrm{asy} \sqrt{\Omega_{m,0}^\mathrm{asy} \, a^{-1} + (1 - \Omega_{m,0}^\mathrm{asy}) \, a^2}},
\end{equation}
allowing to numerically solve Eq. (\ref{eq:H_0_r}) for $H_{T,0}(r)$. Finally, the resulting $H_{T,0}(r)$ and the density parameters in Eq. (\ref{eq:LTB_density_parameters_delta}) are employed to obtain $R(t, r)$ from Eq. (\ref{eq:generalized_Friedmann}). In this way, the radial dependence of all $\Lambda$LTB quantities is fully characterized by setting initial conditions on two functions only, $t_0$ and $\delta(r)$, corresponding to the aforementioned two physical degrees of freedom. 

The result for $R(t, r)$ and the cosmological parameters can be used to numerically integrate the null geodesic equations associated to the LTB metric. We consider the formulation for radially incoming light rays and an observer located at the symmetry center, corresponding to $r = 0$ (see Sec. 3 in \cite{Enqvist:2007vb} for a derivation) \footnote{The hypothesis that the observer is located at the center is motivated by the observed near-perfect isotropy of the CMB. Nevertheless, it is possible to develop models allowing for a certain amount of off-set, e.g. \cite{blomqvist2010supernovae}.}:
\begin{equation} \label{eq:null_geodesics}
    \left\{
    \begin{aligned}
        & \frac{\mathrm{d} t}{\mathrm{d} z} = - \frac{R'(t, r)}{(1 + z) \, \dot{R}'(t, r)} \\
        & \frac{\mathrm{d} r}{\mathrm{d} z} = - \frac{\sqrt{1 - k(r)}}{(1 + z) \, \dot{R}'(t, r)},
    \end{aligned}
  \right.
\end{equation}
where $z$ denotes redshift. This formulation is valid provided that the redshift is a monotonic function of the affine parameter along the geodesics, which we have verified both with the generalized Gaussian and the Oppenheimer-Snyder one.

The geodesic equations yield a numerical solution for $t(z)$ and $r(z)$, which can be used to calculate the generalized scale factor as a function of redshift, corresponding to the angular diameter distance $d_A(z)$ for a $\Lambda$LTB universe:
\begin{equation} \label{eq:R_z}
    R(z) = R(t(z), r(z)) \equiv d_A(z). 
\end{equation}

\subsection{The generalized Gaussian ansatz} \label{sec:GG}
We perform our main analysis with a generalized-Gaussian (GG) set-up. For computational convenience, we do not set an ansatz on $\delta(r)$ directly, but rather on the product $\Omega_{m,0}(r) \, H_0^2(r)$ with respect to its asymptotic value at $r \rightarrow \infty$. We denote this quantity\footnote{K19 instead set $\rho_{m,0}(r) \propto \Omega_{m,0}(r) H_0^2(r)$ and denote this quantity by $\delta(r)$. However, Eq. (\ref{eq:LTB_density_parameters}) with the gauge choice (\ref{eq:gauge}) yields $\Omega_{m,0}(r) H_0^2(r) = \tilde{M}(r)/r^3$. Thus, by noting that $\tilde{M}(r)$ depends on $\rho_{m,0}(r)$ via an integral in Eq. (\ref{eq:mass_fn}), we conclude that $\rho_{m,0}(r)$ and the product $\Omega_{m,0}(r) H_0^2(r)$ are not directly proportional, such that $\delta(r) \neq \epsilon(r)$ (see Eq. (\ref{eq:delta_from_epsilon})).} by $\epsilon(r)$,
\begin{equation} \label{eq:epsilon_def}
    \epsilon (r) \equiv \frac{\Omega_{m, 0}(r) \, H_{T,0}^2(r) - \Omega_{m,0}^\mathrm{asy} \, (H_0^\mathrm{asy})^2}{\Omega_{m,0}^\mathrm{asy}\, (H_0^\mathrm{asy})^2},
\end{equation}
and obtain the following relation with $\delta(r)$ (see Appendix B in \cite{Castello1571650} for a complete derivation):
\begin{equation} \label{eq:delta_from_epsilon}
    \delta(r) = \epsilon(r) + \frac{r}{3} \epsilon'(r).
\end{equation}
This allows to rewrite the density parameters in Eq. (\ref{eq:LTB_density_parameters_delta}) in a form that is easier to implement numerically,
\begin{equation} \label{eq:LTB_density_parameters_epsilon_asy}
   \left\{
    \begin{aligned}
        & \Omega_{m, 0} (r) = \Omega_{m,0}^\mathrm{asy} \left(\frac{H_0^\mathrm{asy}}{H_{T,0}(r)}\right)^2 \, ( 1 + \epsilon(r))  \\ 
        & \Omega_\Lambda(r) = (1 - \Omega_{m,0}^\mathrm{asy}) \left(\frac{H_0^\mathrm{asy}}{H_{T,0}(r)}\right)^2 \\
        & \Omega_{k,0}(r) = 1 - \Omega_{m,0}(r) - \Omega_\Lambda(r).
    \end{aligned}
  \right.
\end{equation}

\begin{figure}[htb!] 
\minipage{0.50\textwidth}
\centering
    \includegraphics[width=1.1\linewidth]{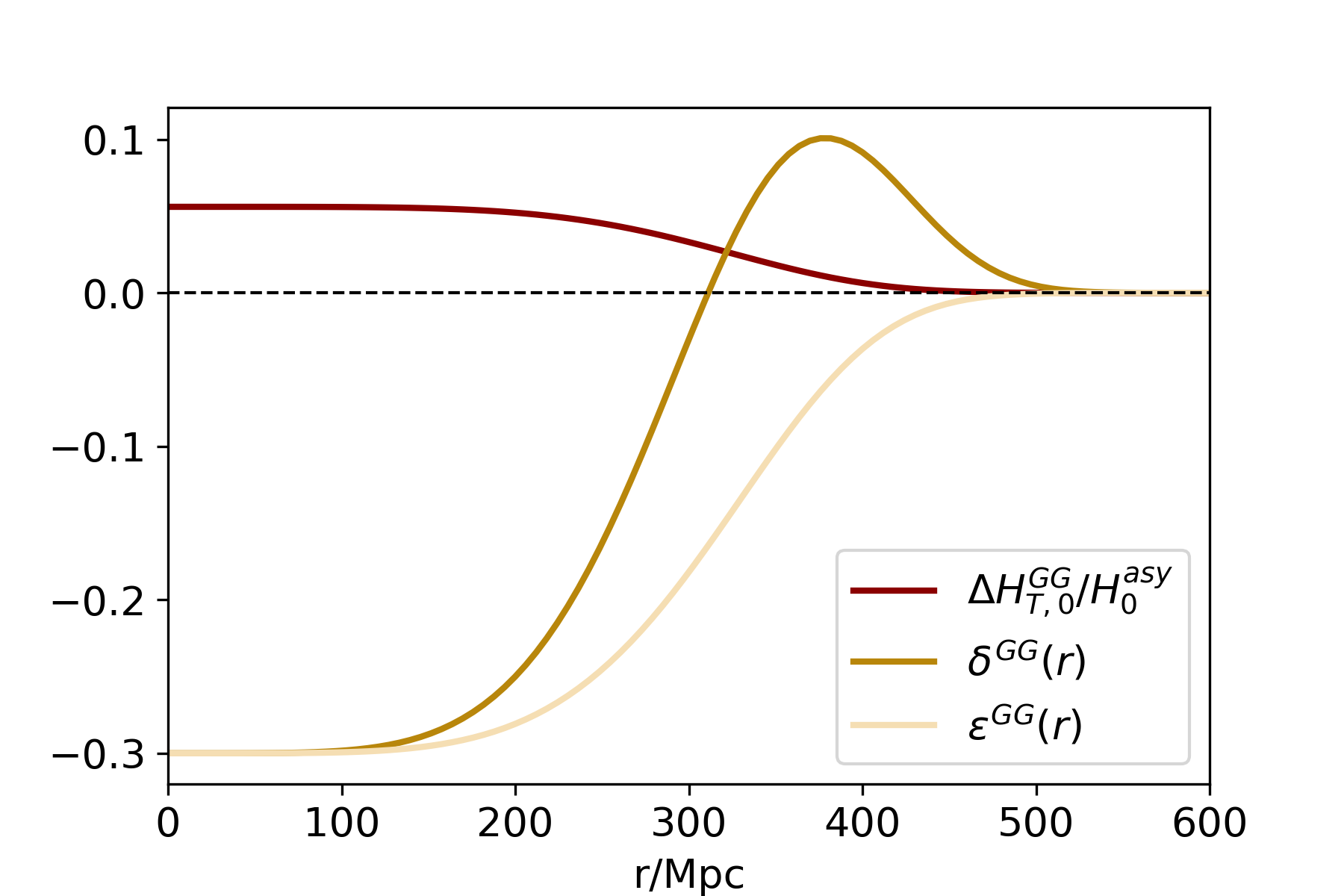}
\endminipage\hfill
\minipage{0.50\textwidth}
\centering
    \includegraphics[width=1.1\linewidth]{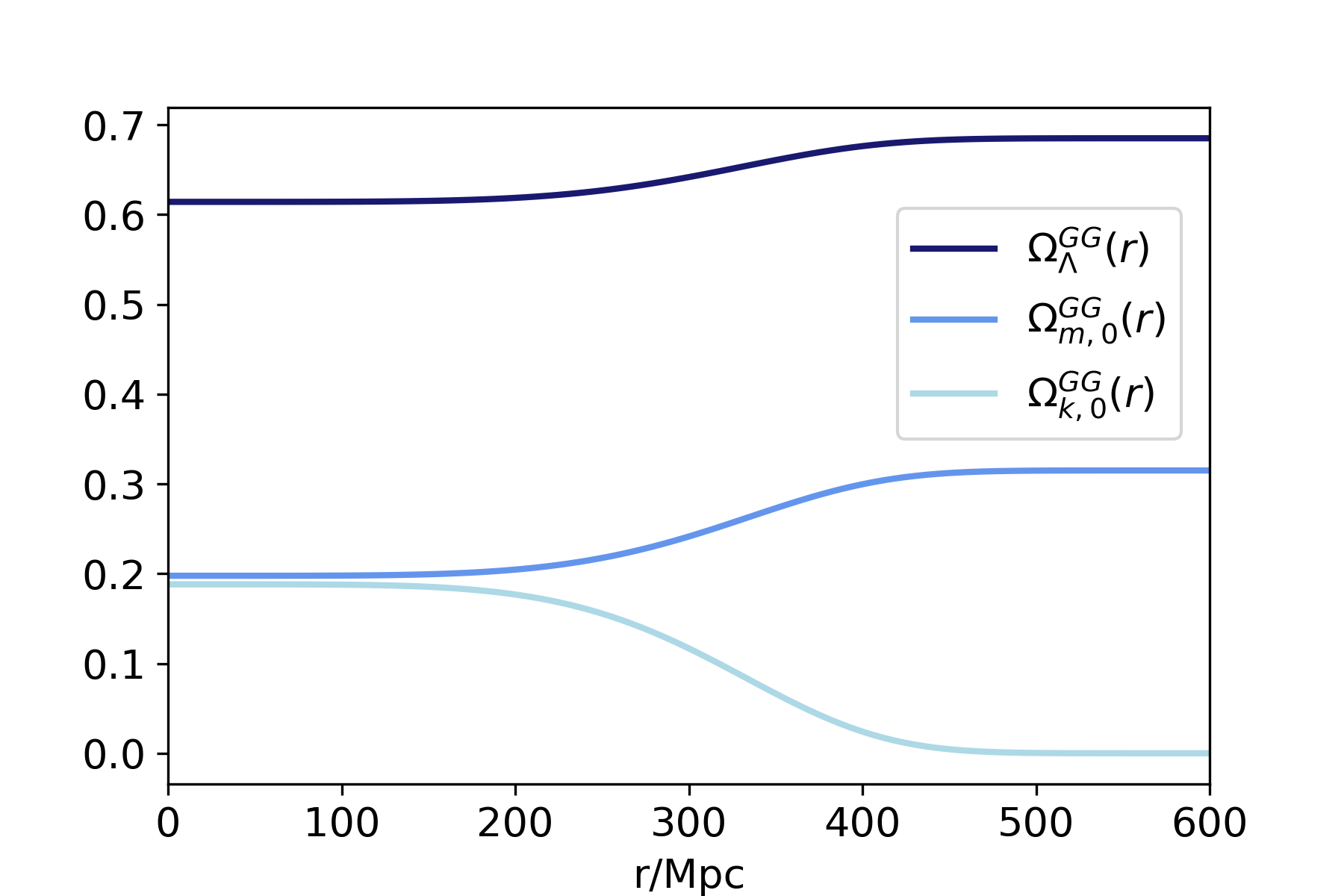}
\endminipage
\caption{Left: The physical matter density contrast $\delta^\mathrm{GG}(r)$ (in brown) and $\epsilon^\mathrm{GG}(r)$ (in pale brown) with $\delta_0 = -0.3$, $\sigma = 300$ Mpc and $n = 5$. The dark red line indicates the predicted variation of $H_{T,0}(r)$ with respect to the \textit{Planck} \cite{Planck:2018vyg} value $H_0^\mathrm{asy} = 67.4$ km s$^{-1}$ Mpc$^{-1}$. Right: The density parameters $\Omega_\Lambda(r)$, $\Omega_{m,0}(r)$ and $\Omega_{k,0}(r)$ (ranging from dark to light blue) obtained with the \textit{Planck} value $\Omega_{m,0}^\mathrm{asy} = 0.315$ and the same parameter configuration as in the left panel.}
\label{fig:GG}
\end{figure}

In our main analysis, we set
\begin{equation} \label{eq:GG_epsilon}
    \epsilon^{\mathrm{GG}}(r) = \delta_0 \,  \mathrm{exp}\left(-\frac{r^n}{2 \sigma^n}\right),
\end{equation}
where $\delta_0$ is the density contrast at $r= 0$, $\sigma$ is the size of the inhomogeneity and the exponent $n$, set to 2 in a standard Gaussian profile, is a free parameter. From Eq. (\ref{eq:delta_from_epsilon}), we obtain
\begin{equation} \label{eq:GG_profile}
    \delta^\mathrm{GG}(r) =  \delta_0 \,  \mathrm{exp}\left(-\frac{r^n}{2 \sigma^n}\right) \left(1 - \frac{n r^n}{6 \sigma^n}\right).
\end{equation}
$\delta^\mathrm{GG}(r)$ and $\epsilon^\mathrm{GG}(r)$, along with the radial dependence of the fundamental cosmological parameters, are shown in Fig. \ref{fig:GG} for the parameter combination $\delta_0 = -0.3$, $\sigma = 300$ Mpc and $n = 5$, roughly corresponding to the prediction for the KBC void \cite{Keenan:2013mfa}.

We note that the shape of $\delta^\mathrm{GG}(r)$ is qualitatively similar to the standard formulations that are employed to fit galaxy data (see e.g. Fig. 3 in \cite{Nadathur:2014qja}), but its mathematical expression is more stable for different choices of parameter values, making it more suitable for our analysis. With this ansatz, all cosmological parameters have a nearly constant behaviour close to the center of the inhomogeneity and then shift to a flat tail, where they attain the corresponding asymptotic values up to machine precision. The radius where the transition to the outer tail takes place is set by $\sigma$, while the steepness is controlled by the parameter $n$, with larger values corresponding to a more abrupt edge.

\section{Numerical investigation} \label{sec:numerical_investigation}
We employ the tools of Bayesian parameter inference \cite{Trotta:2017wnx} to compare the $\Lambda$LTB modelling outlined in Sec. \ref{sec:LTB} with a combination of three data sets: the Pantheon Sample of SNe Ia (both the full catalogue and the standard redshift-binned version of it) \cite{Scolnic:2017caz}, a collection of BAO data points \cite{Beutler2011, Ross:2014qpa, BOSS:2016wmc, Bautista:2017wwp, Zhao:2018gvb} and the CMB distance priors resulting from the latest \textit{Planck} data release \cite{Chen:2018dbv}. The total likelihood is the product of the individual contributions from the data sets, which we write as multivariate Gaussian distributions as specified in Appendix \ref{app:data}. 

We perform a Markov Chain Monte Carlo (MCMC) sampling of the posterior probability distribution for the set of free parameters in our modelling. These are:
\begin{itemize}
    \item The free parameters in the GG ansatz in Eq. (\ref{eq:GG_profile}), $\delta_0$, $\sigma$ and $n$.
    \item The B-band absolute magnitude of the SNe Ia, $M_B$, involved in the calibration of the distances to the SNe according to the procedure from \cite{Dhawan:2020xmp} (see Eq. (\ref{eq:distance_modulus})).
    \item The background cosmology, described by the asymptotic value of the Hubble constant $H_0^\mathrm{asy}$ and the asymptotic matter and baryonic matter density parameters at $t_0$, i.e. $\Omega_{m,0}^\mathrm{asy}$ and $\Omega_{b,0}^\mathrm{asy}$. The first two parameters are involved in the $\Lambda$LTB modelling via Eq. (\ref{eq:LTB_density_parameters_delta}), while the latter enters the formulation of the likelihood for the BAO and CMB data sets (see Eqs. (\ref{eq:c_s}), (\ref{eq:z_d}) and (\ref{eq:z_*})).
\end{itemize}
We set normalized uniform (U) priors on the investigated parameters as specified in Tab. \ref{tab:priors} and allow for underdensities and mild overdensities up to $|\delta_0| = 1$. Since galaxy surveys indicate the potential presence of a local void \cite{Keenan:2013mfa, Whitbourn:2013mwa, Bohringer:2019tyj, Wong:2021fvu}, more pronounced overdensities are disfavoured on the scales of interest, as the results in C21 also indicate.

We separately carry out the MCMC analysis with the redshift-binned and the full version of the Pantheon Sample, thus considering the following combinations:
\begin{enumerate}[label=\Roman*.]
    \item the GG set-up with SNe (redshift-binned) + BAO + CMB; 
    \item the GG set-up with SNe (full) + BAO + CMB.
\end{enumerate}
As a reference analysis, we employ the same data sets (with both SNe configurations) to constrain a flat $\Lambda$CDM cosmology, which can be recovered by setting $\delta_0 = 0$:
\begin{enumerate}[resume*]
   \item $\Lambda$CDM with SNe (redshift-binned) + BAO + CMB; 
   \item the $\Lambda$CDM with SNe (full) + BAO + CMB.
\end{enumerate}
When considering $\Lambda$CDM, the sampled space includes the parameters $M_B$, $H_0$, $\Omega_{m,0}$ and $\Omega_{b,0}$, where we have dropped the labels ``asy" since these quantities do not include a radial dependence in this case.

\renewcommand{\arraystretch}{1.3}
\begin{table} [htb!]
\centering
\begin{tabular}{|c|c|l|}
\hline 
Parameter                              & Model  & $\, \, \, \, \, \, \, \, \, \, \, \, \, \, \, \, \, \, \, \, \,$ Prior\\ 
\hline\hline
$\delta_0$                             & GG  & U[-1, 1]   \\
$\sigma$                     & GG   & U[0, 3500] Mpc   \\
$n$                                     & GG    & U[0, 10] \\
$M_B$                            & All   & U[-30, -10] \\
$H_0^\mathrm{(asy)}$           & All  & U[50, 100] km s$^{-1}$ Mpc$^{-1}$  \\
$\Omega_{m,0}^\mathrm{(asy)}$             & All   & U[0, 1]  \\
$\Omega_{b,0}^\mathrm{(asy)}$              & All   & U[0, 1]  \\
\hline
\end{tabular}
\caption{Investigated parameters with the corresponding models and uniform prior ranges. We note that the upper limit on $\sigma$ is chosen to correspond to a matter density fluctuation enclosing the full SNe sample. The labels ``asy" on the cosmological parameters are put in parentheses since they are not relevant for $\Lambda$CDM.} \label{tab:priors}
\end{table}

All computations were performed with a Python code specifically developed for this work, available on request and based upon the \texttt{emcee} \cite{foreman2013emcee} implementation of the affine invariant ensemble sampling from \cite{GoodmanWeare}. We process the MCMC chains with standard techniques \cite{Sharma2017}, monitoring the state of the sampling through the integrated autocorrelation time and requiring an effective sample size (ESS) of at least 2500 independent samples for each investigated parameter. We visually inspect the chains to check for convergence and remove the burn-in section.

\section{Results and Discussion}\label{sec:results}
\begin{figure} [htb!]
    \centering
    \includegraphics[width=0.8 \linewidth]{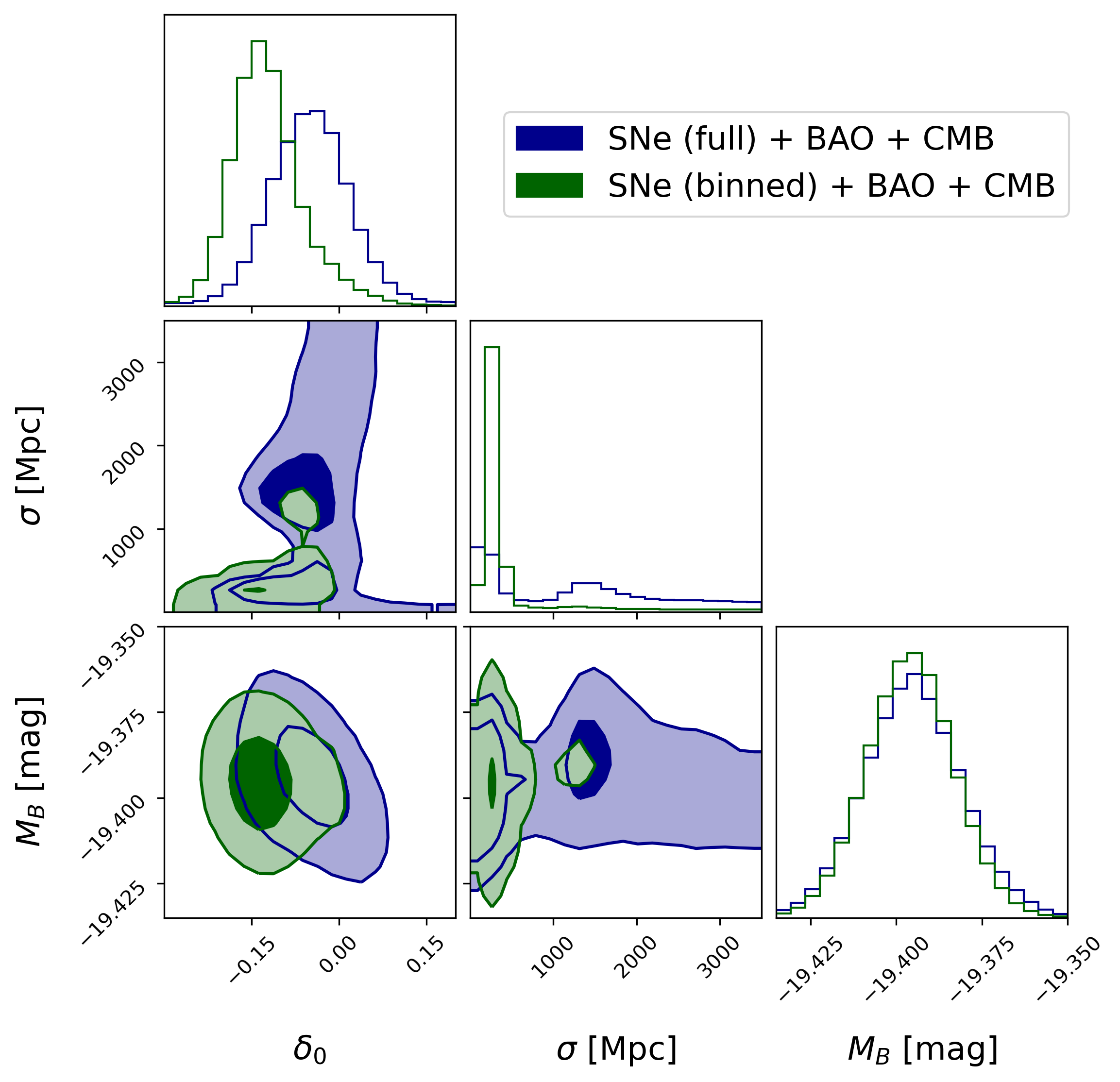}
    \caption{Marginalized one- and two-dimensional constraints for the parameters $\delta_0$, $\sigma$ and $M_B$ obtained with the data set combinations SNe (full) + BAO + CMB (in blue) and SNe (redshift-binned) + BAO + CMB (in green). The dark and light contours in the two-dimensional plots respectively refer to the 1$\sigma$ and 2$\sigma$ levels, corresponding to the 39$\%$ and 86$\%$ credence regions.}
    \label{fig:corner_main}
\end{figure}
We present in Fig. \ref{fig:corner_main} the marginalized one- and two-dimensional posterior distributions obtained for the parameters $\delta_0$, $\sigma$ and $M_B$ in the runs I and II. A complete version of the plot including the full investigated parameter space can be found in Appendix \ref{app:full_corner}. The triangle plots were generated with the \texttt{corner} Python package \cite{corner}. 
\subsection{Constraints on the inhomogeneity} \label{sec:constraints_inhomogeneity}
\begin{table}[htb!]
\centering
\begin{tabular}{|c|c|c|} 
\hline
\multicolumn{1}{|l|}{Parameter} & \multicolumn{1}{l|}{SNe (redshift-binned) + BAO + CMB} & SNe (full) + BAO + CMB                 \\ 
\hline\hline
$\delta_0$       & $-0.13^{+0.06}_{-0.05}$                & $-0.05^{+0.07}_{-0.08}$                                      \\
$\sigma$ [Mpc]                                                           & $300^{+180}_{-60}$  & Bimodal distribution \\
\hline
\end{tabular}
\caption{68$\%$ constraints on the parameters $\delta_0$ and $\sigma$ in the runs I and II.} 
\label{tab:constraints_void}
\end{table}
\noindent
We first focus on the parameters describing the properties of the local matter density profile, whose 68$\%$ constraints are shown in Tab. \ref{tab:constraints_void}. The analysis with the redshift-binned Pantheon Sample yields a preference for a $\simeq -10 \%$ local matter underdensity on scales of approximately 150 to 400 Mpc. In particular, the posterior distribution on $\sigma$ is peaked around a scale of 300 Mpc, which is interestingly close to the predicted size of the KBC void \cite{Keenan:2013mfa}. Our constraint on $\delta_0$, however, corresponds to a shallower inhomogeneity than the $\simeq -30\%$ density drop generally associated with the KBC void \citep{Hoscheit:2018nfl}, but is compatible with other predictions in literature (e.g. \cite{Bohringer:2019tyj}). Moreover, as can be noted from Fig. \ref{fig:corner_GG}, the parameter $n$ is pushed towards large values, corresponding to a inhomogeneity with a steep edge, similar to the OS model presented in Appendix \ref{app:OS}. 

This scenario, however, changes when employing the full Pantheon Sample: the posterior distribution for $\delta_0$ is shifted towards less negative values, such that the 68$\%$ constraints are compatible with the lack of an inhomogeneity (corresponding to $\delta_0 = 0$). This does not highlight a preference for underdensities with respect to overdensities, confirming the findings in C21. Moreover, the parameter $\sigma$ is now characterized by a bimodal distribution, with a first peak located at around 300 Mpc, as in the redshift-binned case, and a second one on larger scales ($\simeq$1500 Mpc). The tail at large $\sigma$ obtained with the binned SNe can be interpreted as a hint of this bimodal behaviour, which however becomes evident only when considering the full data set. This suggests that the binning of the SNe introduces a bias in the constraints, providing a cautionary tale for the use of a compressed version of the Pantheon Sample to constrain inhomogeneous cosmologies.

\subsection[Comparison with $\Lambda$CDM]{Comparison with \boldmath{$\Lambda$}CDM}\label{sec:comparison_LambdaCDM}
We compare the quality of the fit obtained with the GG model and $\Lambda$CDM for both SNe configurations. We employ the Akaike information criterion \cite{Akaike}, which is based upon the value of the indicator
\begin{equation}
     \mathrm{AIC} \equiv -2 \ln \mathcal{L}(\theta_{\mathrm{max}}) + 2k.
\end{equation}
Here, $\mathcal{L}(\theta_{\mathrm{max}})$ is the likelihood evaluated at the set of parameter values maximizing the posterior and $k$ is the number of parameters in the tested model. Given a set of models, the quantity
\begin{equation}
    P_{[i,j]} = \exp{\left(\frac{\mathrm{AIC}_i -\mathrm{AIC}_j}{2}\right)}
\end{equation}
encodes the probability that the $j$-th model minimizes the information loss with respect to a hypothetical ``true" description more than the $i$-th one. 

When employing the binned SNe, we obtain $P_{[\Lambda\mathrm{CDM}, \, \mathrm{GG}]} \simeq 51$, indicating that the GG model is 51 times more likely to be close to the ``true" description of the data than $\Lambda$CDM. Therefore, this result interestingly yields a strong preference for a KBC-sized local inhomogeneity with this data set combination. On the other hand, the analysis with the full SNe gives $P_{[\Lambda\mathrm{CDM}, \, \mathrm{GG}]} \simeq 1$, which does not allow to distinguish between the two models. This outcome is in line with the constraint of $\delta_0 = -0.05^{+0.07}_{-0.08}$ obtained in run II, which is compatible with $\Lambda$CDM ($\delta_0 = 0$). 

\begin{table}[htb!]
\centering
\begin{tabular}{|c|c|c|} 
\hline
\multicolumn{1}{|c|}{Parameter} & \multicolumn{1}{c|}{$\Lambda$LTB (GG)~} & $\Lambda$CDM          \\ 
\hline \hline
$H_0^\mathrm{(asy)}$  [km s$^{-1}$ Mpc$^{-1}$]            & $68.1 \pm 0.5 $          & $67.7 \pm 0.4$     \\
$\Omega_{m,0}^\mathrm{(asy)}$   & $0.311 \pm 0.007$        & $0.312 \pm 0.006$  \\
$\Omega_{b,0}^\mathrm{(asy)}$   & $0.048 \pm 0.001$        & $0.049 \pm 0.001 $   \\
\hline
\end{tabular}
\caption{68$\%$ constraints on the cosmological parameters in the runs II and IV (with the full SNe sample). The labels ``asy" are put in parentheses since they are not relevant for $\Lambda$CDM.}
\label{tab:cosmological_parameters}
\end{table}
We also perform a comparison between the GG model and $\Lambda$CDM in terms of the 68$\%$ constraints on the cosmological parameters, which we present in Tab. \ref{tab:cosmological_parameters} for the combination SNe (full) + BAO + CMB. The credence regions are always compatible between the two models, with only a marginal increase in the error bars in the GG case because of the correlations with the parameters describing the density fluctuation. As pointed out by C21, this indicates that the adoption of an inhomogeneous cosmology does not have an impact on the constraints on the standard cosmological parameters. Fig. \ref{fig:corner_GG} shows that the use of the binned SNe does not change the posterior distributions on the cosmological parameters, such that the considerations above are also valid in this case.

\subsection{Implications for the Hubble tension}\label{sec:Hubble_tension}
We now turn to the key point of this work. As discussed in Sec. \ref{sec:LTB}, in LTB models $H_0$ is a function of the radial coordinate. Therefore, the potential improvement in the Hubble tension with our set-up cannot be assessed by comparing the value of $H_{T,0}$ obtained at $r = 0$ from Eq. (\ref{eq:H_0_r}) with $H_0^\mathrm{asy}$. Instead, a more suitable approach consists in reformulating the tension in terms of a more fundamental parameter: the absolute magnitude of the SNe, here taken in the B-band ($M_B$).

The role of $M_B$ can be understood by considering the SH0ES team's method to determine $H_0$ \cite{Riess:2019cxk}. The first step consists in calibrating the period-luminosity relation of Cepheid variable stars and by employing this relation to estimate the distances to SNe host galaxies that also contain Cepheids. This allows to obtain $M_B$, which can be converted into a value for $H_0$ by considering a data set of SNe in the Hubble flow (e.g. the Pantheon Sample) and by specifying the SNe magnitude-redshift relation within a chosen background cosmology. This implies the Hubble constant is not measured directly and is degenerate with $M_B$, such that a larger value of $M_B$ also leads to an increase in $H_0$. Therefore, as already pointed out in literature (e.g. \cite{Efstathiou:2021ocp}), the Hubble tension can be interpreted in a more fundamental sense as a discrepancy between the larger value of $M_B$ that is inferred from the local distance ladder and the smaller one obtained when also including early-time constraints, such as from the CMB. 

Based upon these considerations, we quantify the level of improvement in the tension by comparing the bounds on $M_B$ obtained for $\Lambda$CDM and the GG model. Provided that the data allows for the existence of an inhomogeneity, an increase in $M_B$ in this second case would lead to an improvement in the tension. However, Tab. \ref{tab:M_B} shows that, for both SNe configurations, the 68$\%$ constraints on $M_B$ in the GG model and $\Lambda$CDM are very close to each other. This result is expected when employing the full SNe, since, as pointed out in Sec.  \ref{sec:constraints_inhomogeneity}, the bounds on $\delta_0$ are consistent with the lack of an inhomogeneity. On the other hand, this outcome is more surprising for the analysis with the redshift-binned SNe, where we observe a strong preference for an underdensity. We interpret this as an indication that, even in a scenario where the data suggest the presence of a local void, the Hubble tension remains unsolved.

\begin{table}[htb!]
\centering
\begin{tabular}{|c|c|c|} 
\hline
Model & SNe (redshift-binned) + BAO + CMB~ & SNe (full) + BAO + CMB  \\ 
\hline \hline
$\Lambda$CDM   & $-19.41 \pm 0.01$                  & $-19.41 \pm 0.01$       \\
GG    & $-19.40 \pm 0.01$                  & $-19.40 \pm 0.01$       \\
\hline
\end{tabular}
\caption{68$\%$ constraints on $M_B$.}
\label{tab:M_B}
\end{table}

\section{Summary and conclusions}
We have approached the so-called void solution to the Hubble tension under more general assumptions than in previous studies \cite{Kenworthy:2019qwq, Hoscheit:2018nfl, Lukovic:2019ryg, Cai:2020tpy}. We have adopted the Lemaître--Tolman--Bondi (LTB) metric to model a spherically-symmetric local matter density fluctuation in a universe containing matter, curvature, radiation and a cosmological constant. Within this formalism, all cosmological parameters acquire a radial dependence and the cosmic evolution is described by a generalized version of the Friedmann equation. In most literature examples, the latter is solved by setting an ansatz on the change of the matter density parameter $\Omega_{m,0}(r)$ (see e.g. \cite{Hoscheit:2018nfl}) or on the variation of the product $\Omega_{m,0}(r) H_0^2(r)$, as in \cite{Kenworthy:2019qwq}. In this study, instead, we express the necessary initial conditions in terms of the matter density contrast $\delta(r)$ and adopt a generalized Gaussian (GG) set-up under the assumption that the Milky Way is located at the center of the inhomogeneity. 

We have constrained the local matter density profile and the background cosmology through a Markov Chain Monte Carlo (MCMC) analysis. We have employed a combination of three data sets: the Pantheon Sample of SNe Ia \cite{Scolnic:2017caz} (both the full catalogue and the redshift-binned version of it), a collection of BAO data points \cite{Beutler2011, Ross:2014qpa, BOSS:2016wmc, Bautista:2017wwp, Zhao:2018gvb} and the distance priors extracted from the latest \textit{Planck} data release \cite{Chen:2018dbv}. The analysis with the redshift-binned SNe yields a preference for a $\simeq 300$ Mpc void (with error bars of the order of 100 Mpc) with depth $\delta_0 = -0.13^{+0.06}_{-0.05}$ at the 68$\%$ level, whose size interestingly matches the prediction for the hypothetical KBC void identified in \cite{Keenan:2013mfa}. In this case, $\delta_0$ is significantly different from zero, suggesting that the GG model provides a better description to the data than $\Lambda$CDM, which is also confirmed by the Akaike information criterion. 

When employing the full Pantheon Sample, however, this scenario changes: we obtain $\delta_0 = -0.05^{+0.07}_{-0.08}$, which is compatible with the lack of an inhomogeneity and agrees with the findings in \cite{Camarena:2021jlr}. Moreover, the posterior distribution for the size of the inhomogeneity ($\sigma$) is bimodal, with a first peak centered at around 300 Mpc and a second one at larger values. We recognize signatures of this bimodal behaviour in the tail at large $\sigma$ values in the posterior distribution obtained with the binned SNe. This suggests that the binning is responsible for suppressing the second peak and we interpret this result as a cautionary tale for the use of a compressed supernova data set to constrain inhomogeneous cosmologies.  

We assess whether our GG model provides a solution to the Hubble tension by reformulating the discrepancy in terms of the (B-band) absolute magnitude of the SNe. This choice is motivated by the fact that the value of $M_B$ is degenerate with the local Hubble constant in a $\Lambda$CDM scenario, such that the observed local increase in $H_0$ is directly correlated to a shift to a higher $M_B$. In our analysis, we obtain compatible bounds on $M_B$ between the GG model and $\Lambda$CDM for both SNe configurations, indicating that the local matter density does not affect the value of $M_B$ and therefore cannot provide an explanation for the tension. This result is particularly interesting in the analysis with the binned SNe, where we observe a strong preference for an underdensity. This conclusion agrees with the claim by \cite{Kenworthy:2019qwq} that the supernova-based determination of the Hubble constant is not impacted by local cosmic structures.

Our results are at least to some extent dependent on the parametrization of $\delta(r)$. The GG set-up we adopt is qualitatively similar to the functional forms that are usually employed to fit galaxy data (see e.g. \cite{Nadathur:2014qja}) and, in addition to $\delta_0$ and $\sigma$, it includes a parameter controlling describing the steepness of the density profile. However, even when observing a marked preference for an underdensity with clear bounds on $\sigma$ and $\delta_0$ (with the redshift-binned SNe), this parameter is poorly constrained. In other words, the data sets that we consider do not provide strong constraints on the shape of the density profile. Nevertheless, from the point of view of the Hubble tension, we believe that a different parametrization would not change our main conclusion that a local underdensity cannot alleviate the discrepancy. This is confirmed by the analysis with a different density profile (based upon an Oppenheimer--Snyder ansatz) presented in Appendix \ref{app:OS}.     

\acknowledgments
We acknowledge the use of the Sunrise HPC facility at Stockholm University and thank Mikica Kocic for technical support. We are grateful to Suhail Dhawan for useful discussions and for providing the matrix with systematic uncertainties for the binned SNe data set. We thank the anonymous referee for valuable comments and are grateful to Martin Sahlén and Erik Zackrisson for their involvement in the master's thesis upon which this work is based. EM acknowledges support from the Swedish Research Council under Dnr VR 2020-03384. 

\newpage
\appendix
\section{Likelihood function and data sets} \label{app:data}
We perform the MCMC analysis with a combination of three data sets: the Pantheon Sample of SNe Ia \cite{Scolnic:2017caz}, a collection of BAO data points \cite{Beutler2011, Ross:2014qpa, BOSS:2016wmc, Bautista:2017wwp, Zhao:2018gvb} and the CMB distance priors from the latest \textit{Planck} data release \cite{Chen:2018dbv}. The total likelihood is the product of the individual contributions from the data sets, in each case corresponding to a $k$-dimensional multivariate normal distribution for a vector of observables $\boldsymbol{X}$,
\begin{equation} \label{eq:likelihood}
    \mathcal{L} = \frac{1}{\sqrt{(2 \pi)^k \mathrm{det(\Sigma)}}} \, \mathrm{exp}\left(-\frac{1}{2} \, \boldsymbol{\Delta}^T  \, \Sigma^{-1} \, \boldsymbol{\Delta} \right).
\end{equation}
Here, we have defined $\boldsymbol{\Delta} \equiv \boldsymbol{X}_{\mathrm{observed}} - \boldsymbol{X}_{\mathrm{theoretical}}$ and $\Sigma$ is the $k \times k$ covariance matrix associated with the data points. In the following, we specify the key features of the three data sets and the nature of $\boldsymbol{X}$ in each case.

\subsection{Type Ia supernovae} \label{app:SNe}
SNe Ia are widely employed as cosmological probes by comparing observations of their apparent magnitude $m$ with the theoretical prediction obtained via the relation
\begin{equation} \label{eq:distance_modulus}
    \mu(z) \equiv m(z) - M = 5 \log_{10} \left(\frac{d_L(z)}{\mathrm{Mpc}}\right) + 25.
\end{equation}
Here, we have introduced the distance modulus $\mu$, which is a function of redshift $z$, while $M$ is the absolute magnitude and $d_L(z)$ is the luminosity distance. In our $\Lambda$LTB model, we consider an observer located at the center of a spherically-symmetric density fluctuation, such that the assumption of isotropy is preserved. Thus, $d_L$ can simply be obtained from Eq. (\ref{eq:R_z}) as
\begin{equation} 
    d_L(z) \equiv (1 + z)^2 \, d_A(z) = (1 + z)^2 \, R(z).
\end{equation}

We compare our $\Lambda$LTB model with the Pantheon Sample of SNe Ia\footnote{Available at \url{https://archive.stsci.edu/hlsps/ps1cosmo/scolnic/}} \cite{Scolnic:2017caz}, consisting of 1048 objects within $0.01 \leq z \leq 2.3$. In the catalog, the B-band apparent magnitude $m_B$ of the SNe is already corrected for stretch, color and the host-galaxy influence and the redshift measurements account for our peculiar motion with respect to the CMB\footnote{In \cite{Scolnic:2017caz}, the transformation of the Pantheon data to the CMB rest frame is performed assuming a FLRW model. However, we expect this to yield a subdominant correction to our results, since the cosmology dependence in the conversion to physical distances is very weak (see Sec. 7.1 in \cite{Carr:2021lcj}). Moreover, the only difference with respect to $\Lambda$LTB models arises from the curvature term \cite{Romano:2012gk}, which is very small for the rather shallow density fluctuations that we consider. We have also explicitly verified that our results do not significantly change when employing the Pantheon data products without peculiar velocities corrections \cite{Carr:2021lcj}, which were released shortly after our analysis was completed.} We consider both the full catalogue and a compressed version of it, binned over the range $0.014 \leq z \leq 1.61$ and including 40 data points.

The B-band absolute magnitude $M_B$ is a free parameter and must be obtained through a calibration of the distance modulus, which is usually performed through observations of Cepheid variable stars. We follow the procedure proposed in \cite{Dhawan:2020xmp}, modelling the Cepheid contribution through the inclusion of an additional data point in the redshift bin $0 \leq z \leq 0.01$. This point is assigned the values $\mu_{\mathrm{Ceph}} = 32.85$, $m_{\mathrm{Ceph}} = 13.613$ and $\sigma_{m_\mathrm{Ceph}} = 0.040$, chosen such that the corresponding $M_B$ and its uncertainty yield the value of $H_0$ obtained in \cite{Riess:2019cxk}. With this set-up, the vector $\boldsymbol{\Delta}$ in Eq. (\ref{eq:likelihood}) is constructed by including $M_B$ in the investigated parameter space and by taking the differences
\begin{itemize}
    \item $m_{\mathrm{Ceph}} - M_B - \mu_{\mathrm{Ceph}}$ for the calibration point;
    \item $m_{\mathrm{observed}} - M_B - \mu_{\mathrm{theoretical}}$ for the data points from the Pantheon Sample.
\end{itemize}

The covariance matrix is calculated as
\begin{equation} \label{eq:covariance_SNe}
    \Sigma_{\mathrm{SNe}} = D_{\mathrm{stat}} + C_{\mathrm{sys}},
\end{equation}
where $D_{\mathrm{stat}}$ is a diagonal matrix containing the square of the statistical errors and $C_\mathrm{syst}$ includes the contribution of systematic uncertainties. For the runs with the redshift-binned version of the catalogue, $C_\mathrm{syst}$ was directly provided by the authors of \cite{Dhawan:2020xmp}, who computed it for the 40 SNe data points and the calibration point by taking 87 sources of uncertainty into account, as discussed in Sec. 2.2 of their paper. For the other runs, we employ the formulation of $C_\mathrm{syst}$ associated with the Pantheon Sample and neglect the covariance with the calibration point, which we expect to play a marginal role.  

\subsection{Baryon acoustic oscillations} \label{app:BAO}
The acoustic oscillations that occurred in the primordial baryon-photon plasma provide a ``standard ruler" for distance measures in cosmology. During the drag epoch, occurring at $z_d \simeq 1060$, the photon pressure was no longer sufficient to prevent the baryons from undergoing gravitational collapse, such that these oscillations got imprinted on the matter distribution. Thus, their scale can be extracted from the observed galaxy power spectrum, yielding a constraint on a quantity in the form $D_X /r_d$, where $D_X$ is a cosmological distance scale and $r_d \equiv r_s(z_d)$ is the sound horizon at the drag epoch. 

We consider a set of BAO data points from 6dGFS \cite{Beutler2011}, SDSS MGS \cite{Ross:2014qpa}, BOSS DR12 \cite{BOSS:2016wmc}, BOSS DR14 \cite{Bautista:2017wwp} and eBOSS QSO \cite{Zhao:2018gvb}, presented in Tab. \ref{tab:BAO} \footnote{Section 3 in \cite{Zumalacarregui2012} shows that the BAO features remain at roughly constant coordinate positions in LTB universes. This justifies the use of BAO data points derived within a FLRW cosmology to perform a fit with a LTB model.}. Overall, we have 10 data points in the effective redshift range $0.106 \leq z^\mathrm{eff} \leq 1.944$, largely overlapping with the Pantheon Sample interval. The covariance matrix is calculated as
\begin{equation} \label{eq:Sigma_BAO}
    \Sigma_{\mathrm{BAO}} = \boldsymbol{\sigma}^T   C  \boldsymbol{\sigma},
\end{equation}
where $C$ is the correlation matrix, given in Tab. \ref{tab:BAO} for each data set, and $\boldsymbol{\sigma}$ is a vector containing the standard deviations associated with the measurements. 

\begin{table}[htb!]
\centering
\begin{tabular}{|c|c|c|cccc|} 
\hline
Data set  & $z_{\mathrm{eff}}$ & Distance measure    & & $10^4C$ & &             \\ 
\hline\hline
6dFGS             & 0.106                                         & $D_V / r_d = 2.976 \pm 0.133$ & - & - & - & -                      \\ 
\hline
SDSS MGS          & 0.15                                          & $D_V / r_d = 4.466 \pm 0.168$ & - & - & - &  -                     \\ 
\hline
                       & 0.38                                          & $D_A / r_d = 10.27 \pm 0.15$  & $10^4$ & 4970 & 1991   &      \\
BOSS DR12        & 0.51                                          & $D_A / r_d = 13.38 \pm 0.18$  & 4970 & $10^4$ & 984  &       \\
                         & 0.61                                          & $D_A / r_d = 15.45 \pm 0.22$  & 1991 & 984 & $10^4$   &      \\ 
\hline
BOSS DR14        & 0.72                                          & $D_V / r_d = 16.08\pm 0.41$   &  - & - & - &  -                        \\ 
\hline
                          & 0.978                                         & $d_A / r_d = 10.7 \pm 1.9$    & $10^4$ & 4656 & 2662 & 248  \\
eBOSS QSO        & 1.23                                          & $d_A / r_d = 12.0\pm 1.1$     & 4656 & $10^4$ & 6130 & 954  \\
                        & 1.526                                         & $d_A / r_d = 11.97 \pm 0.65$  & 2662 & 6130 & $10^4$ & 4257  \\
                        & 1.944                                         & $d_A / r_d = 12.23 \pm 0.99$  & 248 & 954 & 4257 & $10^4$  \\
\hline
\end{tabular}
\caption{BAO data sets considered in our analysis, taken from \cite{Beutler2011, Ross:2014qpa, BOSS:2016wmc, Bautista:2017wwp, Zhao:2018gvb}.} \label{tab:BAO}
\end{table}

Within the $\Lambda$LTB formalism, the distance relations listed in Tab. \ref{tab:BAO} are specified as follows: the angular diameter distance $d_A$ is given in Eq. (\ref{eq:R_z}),
\begin{equation}
    d_A \equiv R(z),
\end{equation}
while the comoving angular diameter distance $D_A$ is
\begin{equation} \label{eq:D_A}
    D_A(z) \equiv (1 + z) \, d_A(z) = (1 + z) R(z)
\end{equation}
and the volume average distance $D_V$ is computed as
\begin{equation} \label{eq:D_V}
    D_V(z) = \left(\frac{D_A^2(z) \, z}{H_R(z)}\right)^{1/3},
\end{equation}
where we underline a dependence on the radial Hubble parameter $H_R(z)$ defined in Eq. (\ref{eq:H_R}) \cite{Zumalacarregui2012}. The scale of the sound horizon at the drag epoch is given by \cite{Efstathiou:1998xx}
\begin{equation} \label{eq:r_d}
    r_d \equiv r_s(z_d) = \int_{z_d}^\infty \frac{c_s \mathrm{d} z}{H_T(z)},
\end{equation}
where the sound speed $c_s$ is expressed as 
\begin{equation} \label{eq:c_s}
    c_s = \frac{c}{\sqrt{3 \left(1 + \frac{3 \omega_b}{4\omega_\gamma}\frac{1}{1+z}\right)}}.
\end{equation}
Here, $c$ is the speed of light, $\omega_\gamma$ is the energy density of the radiation coupling electromagnetically to matter and we have defined $\omega_b \equiv \Omega_{b,0}^\mathrm{asy} (h^\mathrm{asy})^2$, with $h^\mathrm{asy} \equiv H_0^\mathrm{asy}/(100$ km s$^{-1}$ Mpc$^{-1})$. Following \cite{Chen:2018dbv}, we use
\begin{equation}
    \frac{3}{ 4\omega_\gamma} = 31500 \left(\frac{T_\mathrm{CMB}}{2.7 \, \mathrm{K}}\right)^{-4}, \, \, \, \mathrm{with} \, \, \, T_\mathrm{CMB} = 2.7255 \, \mathrm{K}.
\end{equation}
We also introduce $\omega_m \equiv \Omega_{m,0}^\mathrm{asy} (h^\mathrm{asy})^2$ and derive the drag epoch redshift $z_d$ according to the fitting formula proposed in \cite{Hu:1995en},
\begin{equation} \label{eq:z_d}
\begin{split}
    & z_d = 1345 \frac{\omega_m^{0.251}}{1 + 0.659 \, \omega_m^{0.828}} \left[1 + b_1 \omega_b^{b2}\right], \, \, \, \mathrm{with} \\
    & b_1 = 0.313 \, \omega_m^{-0.419} \left[1 + 0.607 \omega_m^{0.674}\right] \, \, \, \mathrm{and} \\
    & b_2 = 0.238 \, \omega_m^{0.223}.
\end{split}
\end{equation}

Because of the high redshifts involved in $r_d$, we have written Eqs. (\ref{eq:c_s})--(\ref{eq:z_d}) in terms of the asymptotic values of the density parameters and $H_0$. For the same reason, when calculating the Hubble parameter $H_T(z)$ entering the integral in Eq. (\ref{eq:r_d}), it is necessary to include the contribution of the radiation density, which becomes non-negligible in this redshift regime. This requires an extension of the standard $\Lambda$LTB formalism described in Sec. \ref{sec:LTB}, where only matter, curvature and a cosmological constant are included. In order to achieve this, we follow the set-up presented in Appendix \ref{app:high_z_H}, which provides a simplified but sufficiently accurate description.

\subsection{Cosmic microwave background} \label{app:CMB}
The so-called CMB distance priors express the information included in the CMB power spectrum in terms of two parameters evaluated at the photon decoupling redshift $z_* \simeq 1090$: the shift parameter $\mathcal{R}$, mainly affecting the height of the peaks in the spectrum, and the acoustic scale $\ell_A$, which controls the spacing between them \cite{WMAP:2008lyn}:
\begin{equation}
    \mathcal{R} = \sqrt{\Omega_{m,0}^\mathrm{asy} \, (H_0^\mathrm{asy})^2} \, D_A(z_*)
\end{equation}
and
\begin{equation}
    \ell_A = \frac{\pi \, D_A(z_*)}{r_s(z_*)}.
\end{equation}
The value of $z_*$ can be calculated through the fitting formula from \cite{Hu:1995en},
\begin{equation}\label{eq:z_*}
    \begin{split}
        & z_* = 1048 \, [1 + 0.00124 \, \omega_b^{-0.738}] [1 + g_1 \, \omega_m^{g_2}], \, \, \, \mathrm{with} \\
        & g_1 = \frac{0.0783 \, \omega_b^{-0.238}}{1 + 39.5 \, \omega_b^{0.763}} \, \, \, \mathrm{and} \\
        & g_2 = \frac{0.560}{1 + 21.1 \, \omega_b^{1.81}}.
    \end{split}
\end{equation}

We employ the constraints on $\mathcal{R}$, $\ell_A$ and $\omega_b \equiv \Omega_{b,0}^\mathrm{asy} (h^\mathrm{asy})^2$ extracted from the latest \textit{Planck} data release \cite{Chen:2018dbv}. We consider the results obtained by assuming a standard FLRW metric, which are applicable within our modelling because the $\Lambda$LTB space-time asymptotically converges to FLRW in the high-redshift regime involved in these measurements. The parameter values with the respective uncertainties and the correlation matrix are presented in Tab. \ref{tab:CMB} and the total covariance matrix is computed as in Eq. (\ref{eq:Sigma_BAO}),
\begin{equation}
    \Sigma_{\mathrm{CMB}} = \boldsymbol{\sigma}^T C \boldsymbol{\sigma}.
\end{equation}

\begin{table} [htb!]
\centering
\begin{tabular}{|c|c|ccc|} 
\hline
\textbf{Parameter} & \textbf{Measured value}   &   & $\mathbf{C}$    &    \\ 
\hline
$\mathcal{R}$      & $1.7502 \pm 0.0046$          & 1.00 & 0.46 & -0.66   \\
$\ell_A$           & $301.471 ^{+0.089}_{-0.090}$ & 0.46 & 1.00 & -0.33   \\
$\omega_b$         & $0.02236 \pm 0.00015$        & -0.66 & -0.33 & 1.00  \\
\hline
\end{tabular}
\caption{Measured values, error bars (corresponding to the 68$\%$ levels) and correlation matrix for the parameters $\mathcal{R}$, $\ell_A$ and $\omega_b$, taken from \cite{Chen:2018dbv}.} \label{tab:CMB}
\end{table}
\noindent

In order to obtain a theoretical prediction for $\mathcal{R}$ and $\ell_A$ within the $\Lambda$LTB formalism, we express the comoving angular diameter distance $D_A$ as in Eq. (\ref{eq:D_A}). The sound horizon at decoupling is given by the same expression as in (\ref{eq:r_d}), this time evaluated at $z_*$:
\begin{equation} \label{eq:r_s_z_*}
    r_s(z_*) = \int_{z_*}^\infty \frac{c_s \mathrm{d} z}{H_T(z)}.
\end{equation}
The set-up to compute the integral is identical to the one presented for the BAO data points and the same considerations concerning the high-redshift regime also apply in this case. 

\section{A simplified set-up for the $\mathbf{\Lambda}$LTB formalism at high redshifts}\label{app:high_z_H}
The $\Lambda$LTB formalism presented in Sec. \ref{sec:LTB} neglects the contribution of radiation. Because of its inhomogeneous pressure component, the inclusion of radiation would make the geodesics of the LTB metric intersect, violating a fundamental assumption involved in our $\Lambda$LTB set-up (see Sec. 2.1 and Appendix A in \cite{Castello1571650} for a complete derivation). Therefore, introducing a nonzero radiation density requires a more complicated framework (see e.g. \cite{Lasky:2006mg}), which is outside of the scope of this work and would anyway not affect the results in the low-redshift regime. 

However, in order to guarantee a realistic description of the BAO and CMB data, it is necessary to include radiation at high redshifts even within our simplified modelling. We achieve this via the approach from \cite{Garcia-Bellido:2008vdn}: first, we identify an effective redshift $z^\mathrm{eff}$ where the curvature component due to the matter density fluctuation has become negligible, but radiation is not relevant yet -- a typical value is $z^\mathrm{eff} \simeq 100$. In this redshift regime, we can approximate the space-time geometry by a FLRW metric. Thus, we impose that the $\Lambda$LTB Hubble parameter at $z^\mathrm{eff}$, which we compute via Eq. (\ref{eq:generalized_Friedmann}), is equal to the value for a flat FLRW universe with an effective Hubble constant $H_0^\mathrm{eff}$ and containing matter and a cosmological constant. Since we are considering a space-time point sufficiently far away from the inhomogeneity, we employ the asymptotic values of the density parameters,
\begin{equation}
\begin{split}
    & (H_0^\mathrm{asy})^2 \, \left[\Omega_{m,0}^\mathrm{asy} \left(\frac{r(z^\mathrm{eff})}{R(z^\mathrm{eff})}\right)^3 + \Omega_{k,0}^\mathrm{asy} \left(\frac{r(z^\mathrm{eff})}{R(z^\mathrm{eff})}\right)^2 + (1 - \Omega_{m,0}^\mathrm{asy} - \Omega_{k,0}^\mathrm{asy}) \right] \\ & \stackrel{!}{=} (H_0^\mathrm{eff})^2 \left[\Omega_{m,0}^\mathrm{asy} \left(1 + z^\mathrm{eff}\right)^3 + (1 - \Omega_{m,0}^\mathrm{asy}) \right].
\end{split}
\end{equation}
We solve this equality for $H_0^\mathrm{eff}$ and insert the result in the Friedmann equation for a flat FLRW universe, where we also include the contribution of radiation. This yields an expression that can be employed as an asymptotic version of Eq. (\ref{eq:generalized_Friedmann}) valid at $z \gtrsim  z^\mathrm{eff}$,
\begin{equation}
    H(z) = (H_0^\mathrm{eff})^2 \, \left[\Omega_{m,0}^\mathrm{asy} \left(1+z\right)^3 + \Omega_{r,0}^\mathrm{asy} \left(1+z\right)^4 + (1 - \Omega_{m,0}^\mathrm{asy} - \Omega_{r,0}^\mathrm{asy}) \right].
\end{equation}
Following \cite{Chen:2018dbv}, we define the radiation density parameter at $t_0$ as
\begin{equation}
    \Omega_{r,0}^\mathrm{asy} = \frac{\Omega_{m,0}^\mathrm{asy}}{1 + z_\mathrm{eq}},
\end{equation}
where the redshift corresponding to the matter-radiation equality is given by 
\begin{equation}
    z_\mathrm{eq} = 2.5 \times 10^4 \, \,  \Omega_{m,0}^\mathrm{asy} \,  \left(\frac{H_0^\mathrm{asy}}{100 \, \, \,  \mathrm{km} \, \, \, \mathrm{s}^{-1} \, \, \, \mathrm{Mpc}^{-1}}\right)^2 \,  \left(\frac{T_\mathrm{CMB}}{2.7 \, \mathrm{K}}\right)^{-4}, \, \, \, \mathrm{with} \, \, \, T_\mathrm{CMB} = 2.7255 \, \mathrm{K}.
\end{equation}
In this way, all high-redshift quantities are fully characterized in terms of the two parameters $\Omega_{m,0}^\mathrm{asy}$ and $H_0^\mathrm{asy}$, which are included in the sampled parameter space in the MCMC analysis.

\section{Analysis with the Oppenheimer--Snyder ansatz} \label{app:OS}
We present below the results obtained by implementing the $\Lambda$LTB set-up with an Oppenheimer--Snyder (OS) ansatz \cite{OSpaper}. In this case, the matter density profile takes the form
\begin{equation} \label{eq:OS_profile}
    \delta^{\mathrm{OS}}(r) = \delta_0 \, \Theta(r_V - r),
\end{equation}
where $\delta_0$ is the density contrast at $r=0$ as in Eq. (\ref{eq:GG_profile}) and $\Theta(r_V - r)$ is the Heaviside step function. This model is characterized by a homogeneous matter distribution for $r < r_V$ and, since the ($\Lambda$)LTB solution at a given radius $r$ only depends on the matter distribution within $r$ (see Eq. (\ref{eq:mass_fn})), this implies that the OS space-time inside the void corresponds to a FLRW metric. 
\begin{figure}[htb!] 
\minipage{0.50\textwidth}
\centering
    \includegraphics[width=1.1\linewidth]{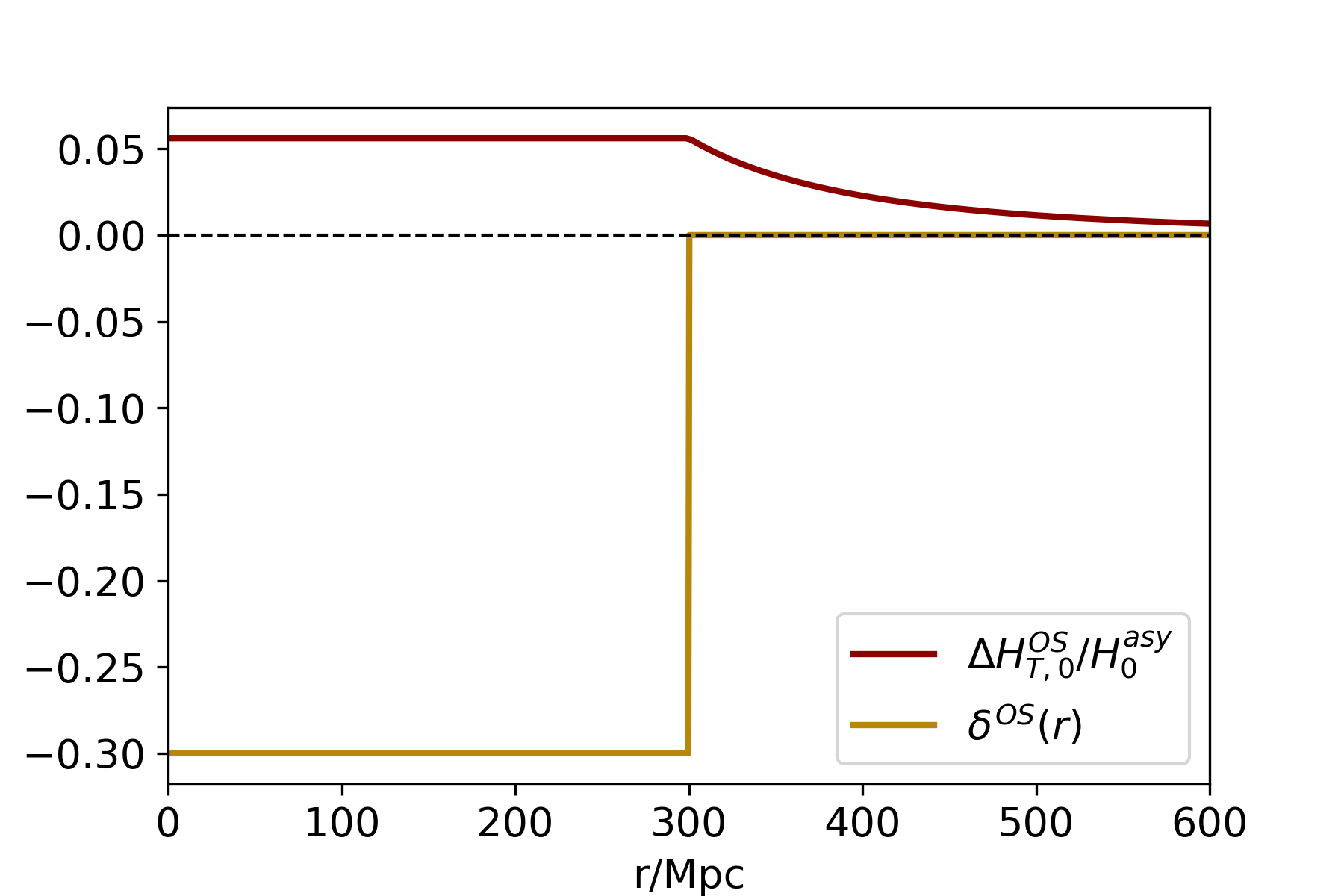}
\endminipage\hfill
\minipage{0.50\textwidth}
\centering
    \includegraphics[width=1.1\linewidth]{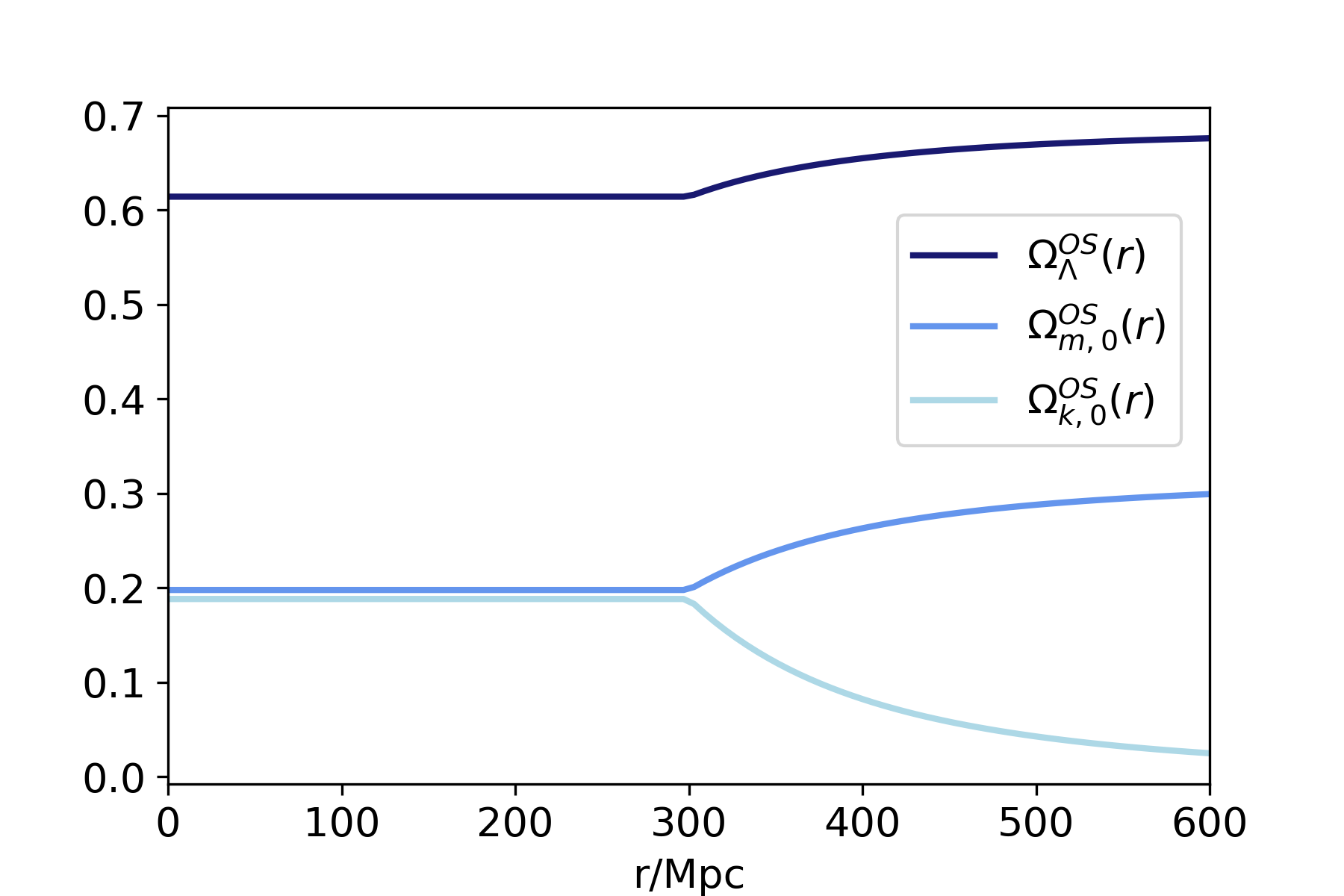}
\endminipage
\caption{Left: The physical matter density contrast $\delta^\mathrm{OS}(r)$ (in brown) and the predicted variation of $H_{T,0}(r)$ with respect to the asymptotic value (in dark red) with $\delta_0 = -0.3$ and $r_V = 300$ Mpc and an asymptotic \textit{Planck} cosmology as in Fig. \ref{fig:GG}. Right: Behaviour of the density parameters $\Omega_\Lambda(r)$, $\Omega_{m,0}(r)$ and $\Omega_{k,0}(r)$ (ranging from dark to light blue) obtained with the same parameter configuration as in the left panel.}
\label{fig:OS}
\end{figure}

From Eq. (\ref{eq:LTB_density_parameters_delta}), we obtain the following exact expressions for the density parameters in the OS model (see Appendix C in \cite{Castello1571650} for a complete derivation),
\begin{equation}
    \Omega_{m,0}(r) =
    \left\{
    \begin{aligned}
    & \Omega_{m,0}^\mathrm{asy} \, \left(\frac{H_0^\mathrm{asy}}{H_{T,0}^\mathrm{OS}(r)}\right)^2 \, (1 - \delta_V) && \text{for $r < r_V$} \\
    & \Omega_{m,0}^\mathrm{asy} \, \left(\frac{H_0^\mathrm{asy}}{H_{T,0}^\mathrm{OS}(r)}\right)^2 \, \left(1 - \delta_V \frac{r_V^3}{r^3}\right) && \text{for $r \geq r_V$},
\end{aligned}
    \right.
\end{equation}
\begin{equation}
    \Omega_\Lambda^\mathrm{OS}(r) = (1 - \Omega_{m,0}^\mathrm{asy}) \left(\frac{H_0^\mathrm{asy}}{H_{T,0}^\mathrm{OS}(r)}\right)^2
\end{equation} 
and
\begin{equation}
  \Omega_{k,0}^\mathrm{OS}(r) = 1 - \left(\frac{H_0^\mathrm{asy}}{H_{T,0}^\mathrm{OS}(r)}\right)^2 \left \{
  \begin{aligned}
    & (1 - \delta_V \, \Omega_{m,0}^\mathrm{asy}) && \text{for $r < r_V$} \\
    &  \left(1 - \delta_V \, \, \Omega_{m,0}^\mathrm{asy} \frac{r_V^3}{r^3}\right) && \text{for $r \geq r_V$}. 
  \end{aligned}
  \right.
\end{equation} 
The radial behaviour of the fundamental cosmological parameters is portrayed in Fig. \ref{fig:OS} for the parameter combination $\delta_0 = -0.3$ and $r_V = 300$ Mpc, mimicking the KBC void \cite{Keenan:2013mfa}. All parameters are constant for $r < r_V$, as expected for a FLRW metric, while a radial dependence appears for $r > r_V$ as a result of the density transition at $r \equiv r_V$.

\begin{figure} [htb!]
    \centering
    \includegraphics[width=0.9\linewidth]{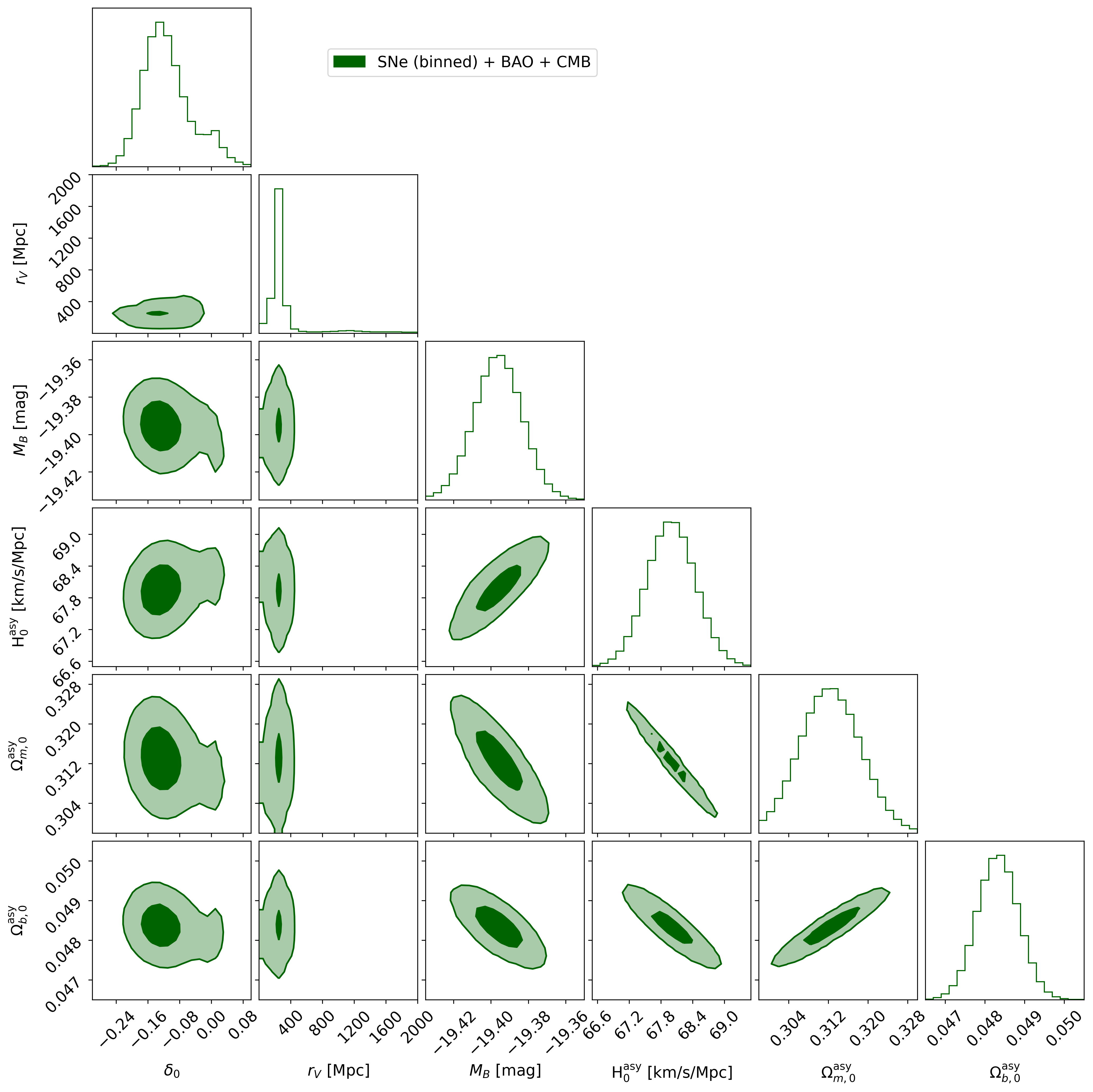}
    \caption{Marginalized one- and two-dimensional probability density functions for the analysis with the OS ansatz and the data set combination SNe (redshift-binned) + BAO + CMB. The dark and light contours in the two-dimensional plots respectively refer to the 1$\sigma$ and 2$\sigma$ levels, corresponding to the 39$\%$ and 86$\%$ credence regions.}
    \label{fig:corner_OS}
\end{figure}

We perform the MCMC sampling for the parameter space $\{\delta_0, \, r_V, \, M_B, \, H_0^\mathrm{asy}, \, \Omega_{m,0}^\mathrm{asy}, \, \Omega_{b,0}^\mathrm{asy}\}$ with the same uniform prior ranges as in Tab. \ref{tab:priors} for the listed parameters and U[10, 5500] Mpc\footnote{The upper bound on $r_V$ is higher than the one on $\sigma$, since the latter does not correspond to the radius of the inhomogeneity (as $r_V$ does), but only measures its size up to the middle of the transition to the outer density bump (see Fig. \ref{fig:GG}). Moreover, we do not set a lower bound of 0 Mpc on $r_V$ as this would lead to numerical instabilities. Nevertheless, since the first supernova data point is located further away than 10 Mpc, this does not affect the results of the analysis.} for $r_V$. In this case, we only employ the data set combination SNe (redshift-binned) + BAO + CMB, yielding the marginalized one- and two-dimensional posterior distributions in Fig. \ref{fig:corner_OS}.

We observe a preference for a local underdensity with 68$\%$ credence regions $\delta_0 = - 0.12^{+0.07}_{-0.05}$ and $r_V = 260^{+280}_{-60}$ Mpc, which are in agreement with the constraints for the GG model with binned SNe (see Tab. \ref{tab:constraints_void}). As with the GG model, the posterior distribution for $r_V$ has a tail at large values, which can be interpreted as a hint of the bimodal distribution that would be obtained with the full SNe sample. The bounds on the cosmological parameters and $M_B$ are compatible with the GG results too, leading to the same considerations as in Secs. \ref{sec:comparison_LambdaCDM}--\ref{sec:Hubble_tension}. In this case, we obtain $P_{[\Lambda\mathrm{CDM}, \, \mathrm{OS}]} \simeq 54$, indicating that the OS model is 54 times more likely than $\Lambda$CDM to be close to the ``true" description of the data set configuration SNe (binned) + BAO + CMB. 
\section{Full corner plot for the GG model} \label{app:full_corner}
\begin{figure} [htb!]
    \centering
    \includegraphics[width=1.0 \linewidth]{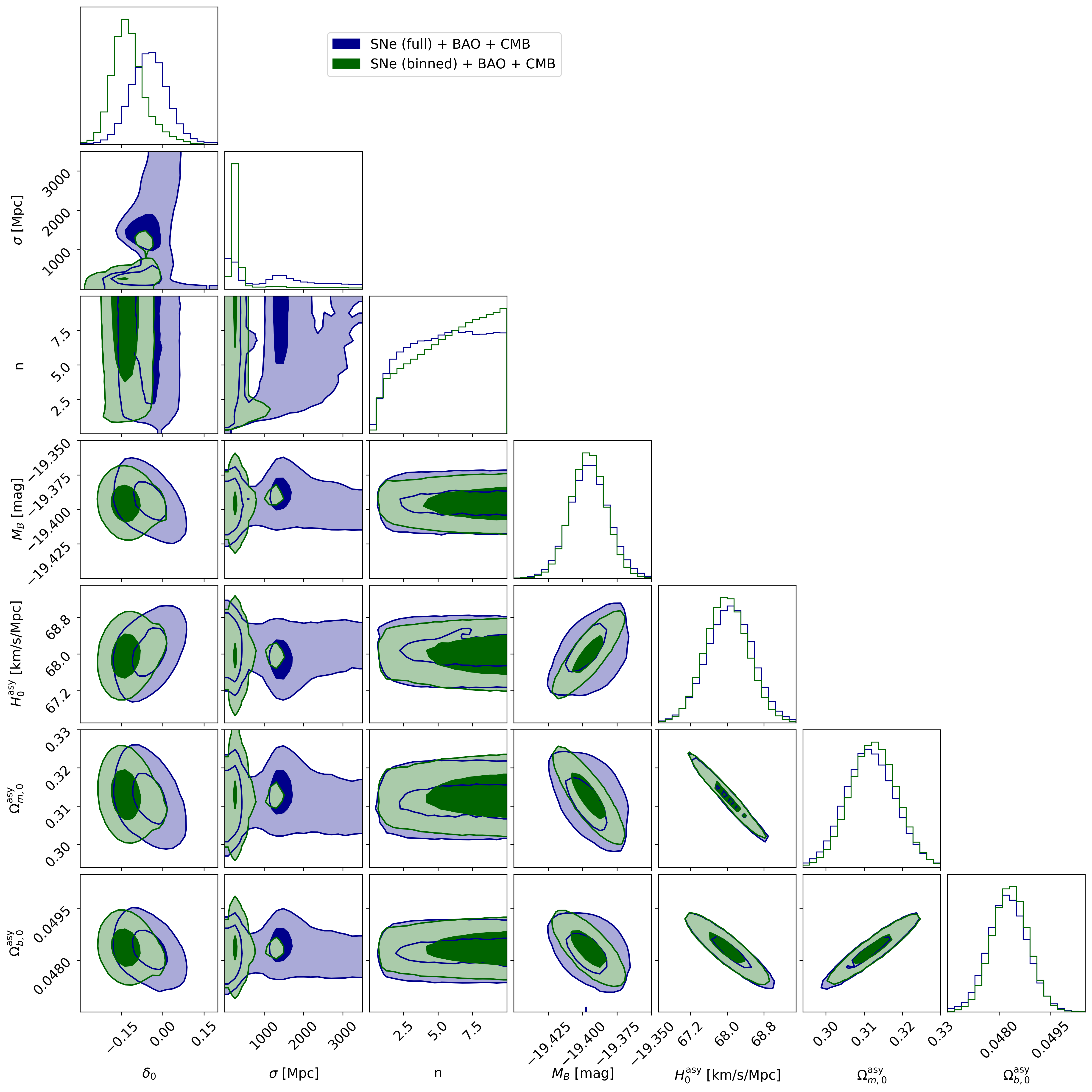}
    \caption{Marginalized one- and two-dimensional probability density functions for the analysis with the GG ansatz, with SNe (full) + BAO + CMB (in blue) and SNe (redshift-binned) + BAO + CMB (in green). The dark and light contours in the two-dimensional plots respectively refer to the 1$\sigma$ and 2$\sigma$ levels, corresponding to the 39$\%$ and 86$\%$ credence regions.}
    \label{fig:corner_GG}
\end{figure}
Fig. \ref{fig:corner_GG} shows the marginalized one- and two-dimensional posterior distributions obtained for the full parameter space in the analyses with SNe (full) + BAO + CMB (in blue) and SNe (redshift-binned) + BAO + CMB (in green). The results for $\delta_0$ and $\sigma$ obtained with the redshift-binned SNe highlight a preference for a underdensity with a $\simeq$ 300 Mpc size, while the parameter $n$ is pushed towards large values within the allowed range. The use of the full SNe data set instead shifts the distribution for $\delta_0$ towards less negative values, such that the $68 \%$ constraints are compatible with $\Lambda$CDM ($\delta_0 = 0$). The distribution for $\sigma$ is in this case bimodal, with a first peak at $\simeq 300$ Mpc and a second one at $\simeq 1500$ Mpc, while the distribution for $n$ reaches a plateau towards large values. On the other hand, the constraints on $M_B$ and on the cosmological parameters are very similar for the two SNe configurations and are compatible with the results obtained in the analysis with $\Lambda$CDM.

\bibliographystyle{JHEP}
\bibliography{biblio}

\end{document}